\RequirePackage{fixltx2e}
\documentclass[nofootinbib,reprint,noeprint,  prc,aps,10pt,superscriptaddress,floatfix]{revtex4-1}

\usepackage{graphicx}
\usepackage{hyperref}
\usepackage{epstopdf}
\usepackage{amsmath}
\usepackage{amssymb}
\usepackage[amsmath,thmmarks]{ntheorem}
\usepackage{bm}
\usepackage{cleveref}
\usepackage{subfigure}
\usepackage{epstopdf}
\usepackage{array}
\usepackage{ifthen}
\usepackage{booktabs}
\usepackage{textgreek}
\usepackage{multirow}
\usepackage{enumitem}
\usepackage{dcolumn}
\usepackage{comment}
\newcolumntype{C}[1]{>{\centering\let\newline\\\arraybackslash\hspace{0pt}}m{#1}}

\let\oldtabular\tabular
\renewcommand{\tabular}{\large\oldtabular}

\usepackage{color}
 
\graphicspath{{}}

\begin{document}

\title{Solid state breakdown counter for magnetic monopole and other highly ionizing particles searches}

\newcommand{\Stanford}{\affiliation{Physics Department, Stanford University, Stanford CA, USA}}
\newcommand{\Alberta}{\affiliation{Physics Department, University of Alberta, Edmonton Alberta, Canada}}

\author{I.~Ostrovskiy}\email[Corresponding author: ]{ostrov@stanford.edu}\Stanford


\author{J.L.~Pinfold}\Alberta

\date{\today}

\begin{abstract}
Solid state breakdown counters combine threshold properties of nuclear track detectors with the convenience of electronic event registration in real time. Here we discuss the principle of operation of the breakdown counters and their potential applicability to the existing (MoEDAL) and foreseen (CosmicMoEDAL) searches of magnetic monopoles and other highly ionizing rare particles. Avenues of R\&D that may lead to an increased appeal of the technology are also described.
\end{abstract}


\maketitle

\section{Introduction}\label{sec:Introduction} 
The magnetic monopole is a hypothetical particle that was postulated to exist by Dirac to explain the apparent quantization of the electric charge~\cite{Dirac:1931}. Different monopole models exist today (see~\cite{Rajantie:2012,Milton:2006} for recent reviews). For example, as was shown by G.'t Hooft, any Grand Unified Theory (GUT) that incorporates electromagnetism can contain magnetic monopoles~\cite{Hooft:1974}. Heavy monopoles with a mass comparable to the GUT scale (\(\sim10^{16}\) GeV) could have been produced in the early universe during a symmetry-breaking phase transition. They are expected to have large ionization losses, thousands times larger than that of a minimally ionizing particle. Intermediate mass monopoles (IMM) exist in theories with several stages of symmetry breaking. IMM are also heavily ionizing. An upper constraint on the flux of the monopoles in our galaxy (\(\sim10^{-15} cm^{-2}s^{-1}sr^{-1}\)) was suggested by Parker~\cite{Parker:1970} based on the observed values of the galactic magnetic field and the fact that the monopoles would be accelerated by the field, hence draining its energy. Monopoles with low enough masses could be produced at particle accelerators, although predicting the rate and kinematics of monopole production is difficult because perturbative field theory does not apply. 

The current strongest limits on monopole fluxes in the cosmic rays come from MACRO~\cite{Macro:2002} for monopoles with masses \(\sim10^{10}-10^{18}\)GeV, SLIM~\cite{Slim:2008} for IMM with masses down to \(\sim10^{5}\) GeV, RICE~\cite{Rice:2008} and ANITA-II~\cite{Anita:2011} for ultra-relativistic monopoles. The planned next generation experiment, Cosmic-MoEDAL, aims to have 50-100 times larger detection area than MACRO and SLIM, improving the sensitivity to monopoles with a wide range of masses \(\sim10^5-10^{18}\)GeV and reaching for the first time the Parker bound for masses \(\sim10^{5}-10^{9}\)GeV~\cite{CosmicMoEDAL:2014}.

Magnetic monopoles with even lower masses are searched for at different particle accelerators every time a higher energy beam becomes available (for a summary, see~\cite{Moedal:2014}). The monopole search at the LHC will be conducted by a dedicated experiment, MoEDAL, which was approved by CERN to begin data taking in 2015. 
Apart from the primary goal of magnetic monopole searches, the MoEDAL's physical program includes searches for dyons (particles with both magnetic and electric charges), nuclearites (hypothetical highly ionizing aggregates of u, d, and s quarks), Q-balls (super-symmetric coherent states of squarks, sleptons, and Higgs fields), and other highly ionizing avatars of new physics~\cite{Moedal:2014}.

Low expected fluxes of cosmic magnetic monopoles call for experiments covering very large surface areas, necessitating simple and low cost detector designs. Very high ionization energy losses, typically expected in monopole models, suggest detectors with high dE/dx thresholds to facilitate discrimination of backgrounds, which are especially abundant in accelerator experiments.
Below we describe a detection technique of solid state breakdown counters (SSBC) that may satisfy both these conditions while offering additional advantages over the currently used nuclear track detectors (NTDs). 

\section{Principle of operation and design}\label{sec:Design} 
\begin{figure}[htbp]
 \centering
\includegraphics[width=0.5\textwidth]{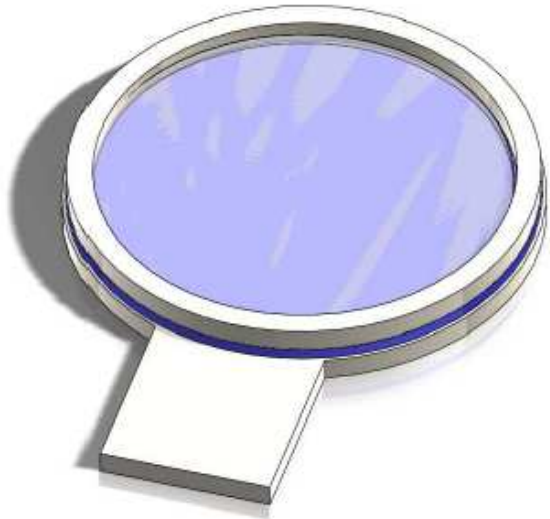}
  \caption{Schematic diagram of a MOM-type SSBC. The design consists of a metallic substrate (anode, shown in grey), a thin film dielectric (a few \(10^1\)nm) layer, shown in deep blue, and a top metal electrode (a few \(10^1\) nm), shown in light blue. The top-most grey metal ring facilitates ohmic contact with the top electrode. Example materials include aluminum and titanium for the electrodes, and their corresponding oxides for the dielectric. Square shape can be used and may be more appropriate for applications requiring covering large surface areas.}
 \label{fig:design}
\end{figure}

The SSBC is a simple technology that dates back almost 50 years~\cite{Klein:1975}. The operating principle of the SSBC is based on an electric breakdown of a thin film capacitor initiated by a highly ionizing particle passing through the sensitive region. 
The design of the existing SSBC utilizes a metal on semiconductor (MOS) structure (see, for example, devices used in~\cite{Smirnov:2012}). For the monopole searches, however, more advantageous could be another type of SSBC  based on a metal on metal (MOM) structure. While practically unknown, the MOM SSBC has simpler design and more stable operation in the presence of low ionizing background radiation~\cite{pte:1998}, which makes it particularly well suited for collider searches. Figure~\ref{fig:design} shows a schematic view of the MOM-type SSBC.

The counter is operated below the intrinsic breakdown voltage. 
The breakdown of the SSBC initiated by a highly ionizing particle does not lead to the short-circuit, because the energy released in the process leads to immediate sublimation of a small portion of the top electrode (cathode). Destruction of the small part of the cathode isolates a microscopic shorted region, of the order of a few square micrometers in size, that allows the detector to continue operation. In high rate applications, the gradual loss of the sensitive surface with running time may need to be taken into account, but for rare event searches the effect is inconsequential. One can note that to enable this mechanism, the top metal electrode needs to be thin enough to allow sublimation, given the energy released in the breakdown. If a reasonably large (0.5-10 k\(\Omega\)) resistor is used in series with the device, and ignoring the heat loss due to the short time scale of the breakdown (\(\sim10^{-9}\) s~\cite{pte:1998}), the released energy \(E\) is related to the capacitance of the device:
\begin{equation}
E = \frac{1}{2}\epsilon\frac{A}{d}(V^2_{bd}-V^2_{res}),
\end{equation}
where \(\epsilon\), $A$, and $d$ are the absolute permittivity, the area, and thickness of the oxide layer,  respectively, $V_{bd}$ is the breakdown voltage, and $V_{res}$ is the residual voltage immediately after breakdown. In practice, this translates into required thicknesses of the cathode electrode on the order of a fraction of a micron or less. Typical pulse heights of SSBC signals are from a fraction of a volt to a few volts. A critical aspect of the SSBC technology is the quality of the oxide film. A simple and low cost anodizing process has been used to produce high quality films~\cite{pte:1998}.

Only a highly ionizing particle can trigger the breakdown at operating voltages, but little work has been done so far to systematically quantify the threshold dependence on various design choices, such as material, thickness, and doping of the oxide layer. It has been shown that the threshold stopping power depends on the operating voltage~\cite{Donichkin:1979}, allowing one to fine-tune the discrimination ability for a given device, or create a stack of devices with varied thresholds by operating individual devices at different voltages. The existing devices are optimized to be triggered only by heavy ions, e.g. fission fragments, while being completely insensitive to accompanying high fluxes of alpha, beta, neutron and gamma radiation. Threshold ionization losses of 10~\cite{Eismont:1995} to 70~\cite{Donichkin:1979} MeV$\cdot$cm$^2\cdot$mg$^{-1}$ are sometimes quoted for the MOS designs, depending on the voltage and device. A similar value was reported for the MOM detector~\cite{pte:1998}. When exposed to a constant flux of a given type of the highly ionizing radiation, the dependence of breakdown rates on the voltage exhibits the typical ``rising until plateau'' trend. The absolute efficiency of registration at the plateau is very high, with typical devices having at least 80\% reported absolute efficiency. Several basic characteristics of the SSBC are listed in Table~\ref{tab:props}. 

\begin{table*}[t]
\centering
\begin{tabular}{l@{\hskip 0.75in} c@{\hskip 0.75in}}
\hline\hline
Operating voltage, V & \bf{1-100} \\
Signal amplitude, V & \bf{0.05 - 5} \\
Detection efficiency at plateau, \% & \bf{$>$80\%} \\
Detection thresholds, MeV$\cdot$cm$^2\cdot$mg$^{-1}$ & \bf{10-100}  \\
Operational resource, breakdowns$\cdot$cm$^{-2}$ &  \bf{$\sim$1$\cdot$10$^6$ }\\
\hline\hline
\end{tabular}
  \caption{Basic properties of solid state breakdown counters (indicative values).}
\label{tab:props}
\end{table*}

\section{Applicability to searches for magnetic monopole and other highly ionizing particles}\label{sec:Monopoles}
Currently, the plastic nuclear track detector (NTD) is a staple detection technique in many monopole searches~\cite{Pinfold:2009}. Both MoEDAL and Cosmic-MoEDAL experiments contain the NTD as a core detection technology. An apparent disadvantage of NTDs is that they require chemical etching and optical scanning to extract the signal. In principle, the SSBC naturally combines the advantages of the NTD with the convenience of electronic event registration in real time. 
Existing SSBC devices already demonstrate high triggering thresholds~\cite{Eismont:1995,Donichkin:1979,pte:1998}, similar to that of the NTDs. The SSBC is operated at moderate voltages (\(\sim\)1-100 V) and outputs \(\sim\)0.05-5 V signals, making the event registration straightforward.
At present, however, a single SSBC provides only on/off detection, without quantitative information about the particle's dE/dx. This limitation can be overcome by using a stack of devices with different calibrated thresholds. 

The straightforward design of the SSBC, especially in the case of the MOM variant, makes it easy to produce large quantities of devices using inexpensive materials and a well established technology. Due to the high thresholds, the devices are insensitive to most background sources, relaxing requirements for material selection, manufacturing standards, and operational precautions. Individual devices with a few cm$^2$ readout areas have been produced and operated, based on existing applications. While there does not seem to be an upper limit on the number of individual devices that could be chained together in a single readout channel, larger areas of individual devices should be achievable with dedicated R\&D  (a potential limiting factor could be the increasing difficulty to maintain high quality of the thin film oxide layer). 

Overall, the SSBC technology is likely to provide an attractive alternative, or a complement, to the NTDs, but some additional R\&D may be needed to realize its full potential.

\section{Avenues for additional improvements}
While the existing SSBC devices already have thresholds similar to the NTDs, there is a realistic potential to further control the minimum specific ionizing losses required to trigger the SSBC. The dependence of the threshold on operating voltage has already been demonstrated, with lower voltages requiring particles with higher stopping power to trigger the breakdown. Additional  handles are possible with the choice of the material and the thickness of the oxide layer, as well as doping of the layer by specific elements. Apart from increasing or decreasing the threshold of individual devices, another goal of the R\&D could be to create an array of devices with different thresholds - stacks of SSBCs. Such stacks could be, in principle, used to determine the actual ionizing loss of a passing highly ionizing particle. This would unambiguously confirm the characteristic signature trend of the monopole energy loss with energy, which is opposite to the one of an electrically charged particle. 
It is also worth investigating whether the commercially available MOM capacitors could be re-purposed as breakdown detectors. Even if out-of-the-box capacitors are not workable as a SSBC, collaboration with the industry can result in a very inexpensive large scale solution due to the established manufacturing capabilities of commercial companies. 
 
\section{Summary and outlook}
A technique for detection of highly ionizing particles using solid state breakdown counters is presented. The SSBC exhibits high registration thresholds, similar to that of the nuclear track detectors. In addition, the existing SSBC technology provides the convenience of electronic registration and simplicity of fabrication, operation, and signal extraction. These properties make it a viable candidate for magnetic monopole and other highly ionizing rare particle searches. With additional R\&D, discussed in the text, the SSBC may become an attractive technology for the detection of highly ionizing particles in addition to NTDs, which are currently a detector of choice by such experiments as MoEDAL and CosmicMoEDAL. The MoEDAL collaboration has expressed interest in the further investigation of the SSBC technology. The primary tasks that are foreseen in the context of this experiment are the following: 1) modeling of the detector; 2) creation of a working prototype of an individual SSBC readout channel; 3) testing of the prototype in a heavy-ion beam; 4) demonstration of the satisfactory background rejection by exposure of a test detector to LHC backgrounds as part of the MoEDAL detector and scalability to large areas.



\bibliography{ssbc}

\end{document}